\documentclass[12pt]{article}
\usepackage{graphics, color}
\usepackage{graphicx}
\usepackage{amssymb}

\newcommand{\sect}[1]{\section{#1}\setcounter{equation}{0}}

\def\gsim{\, \rlap{$>$}{\lower 1.1ex\hbox{$\sim$}}\,}
\def\lsim{\, \rlap{$<$}{\lower 1.1ex\hbox{$\sim$}}\,}

\begin{document}


\begin{titlepage}
\bigskip
\bigskip\bigskip\bigskip
\centerline{\Large \bf  Cosmic String Structure}
\smallskip
\centerline{\Large \bf  at the Gravitational Radiation Scale}
\bigskip\bigskip\bigskip

\centerline{\large Joseph Polchinski}
\bigskip
\centerline{\em Kavli Institute for Theoretical Physics}
\centerline{\em University of California}
\centerline{\em Santa Barbara, CA 93106-4030} \centerline{\tt joep@kitp.ucsb.edu}
\bigskip
\bigskip
\centerline{\large Jorge V. Rocha}
\bigskip
\centerline{\em Department of Physics}
\centerline{\em University of California}
\centerline{\em Santa Barbara, CA 93106} \centerline{\tt jrocha@physics.ucsb.edu}
\bigskip
\bigskip
\bigskip\bigskip


\begin{abstract}
We use our model of the small scale structure on cosmic strings to develop further the result of Siemens, Olum, and Vilenkin that the gravitational radiation length scale on cosmic strings is smaller than the previously assumed $\Gamma G\mu t$.  We discuss some of the properties of cosmic string loops at this cutoff scale, and we argue that recent network simulations point to two populations of cosmic string loops, one near the horizon scale and one near the gravitational radiation cutoff.
\end{abstract}
\end{titlepage}
\baselineskip = 17pt

\sect{Introduction}

Cosmic strings and their dynamics have been studied for more than twenty-five years.  Although much is known about the formation and evolution of cosmic string networks~\cite{VilShell, Hindmarsh:1994re}, the complexity of these systems still stands in the way of a complete understanding.  It is generally believed that, after their formation, the networks evolve into a scaling regime where all length scales grow linearly with the Hubble time.  This would imply that  the typical size of cosmic string loops is a fixed fraction $\alpha$ of the horizon scale.\footnote{There is not complete agreement even on this point: see Ref.~\cite{Bevis:2006mj} and references therein.}  However, there is no consensus on the value of the proportionality constant  $\alpha$; in fact, estimates range over tens of orders of magnitude.

A key question is whether the smoothing of the string by gravitational radiation is necessary for scaling of the loop sizes.   If so,  $\alpha$ will depend on the dimensionless string tension $G\mu$, because the emission of gravitational radiation is proportional to this parameter.  If not, then the evolution of the network will be purely geometric, and $\alpha$ will be a pure number, independent of $G\mu$.  For many years, simulations appeared to show loops forming at the short distance cutoff scale, and this was interpreted as implying the need to include gravitational radiation.  Several recent simulations, however, indicate that the loops are actually forming above the cutoff scale, so that $\alpha$ will be independent of $G\mu$. 
Refs.~\cite{Vanchurin:2005pa,Olum:2006ix} finds notably large loops, $\alpha \sim 0.1$, while Refs.~\cite{Ringeval:2005kr,Martins:2005es}  find loops a few orders of magnitude smaller.

In Ref.~\cite{Polchinski:2006ee} we attempted to construct an analytic model of the small scales in cosmic string networks.  We had some success in understanding the two-point functions of the tangent vector and velocity on long strings.  However, loop production was found to diverge at small scales, and we had no means to deal with the resulting strongly nonlinear dynamics.

In Sec.~2 we use our model to study gravitational radiation from long strings.  It was long assumed that this radiation smoothed the strings on scales of order the horizon length times $G\mu$, so this would be the size of loops if they formed at the gravitational radiation cutoff~\cite{Bennett:1989ak,Quashnock:1990wv}.  However,
Ref.~\cite{Siemens:2001dx} showed the radiation from long strings had been overestimated, and so it becomes important only at a shorter length scale, proportional to a larger power of $G\mu$~\cite{Siemens:2002dj}; the exponent depends on the power spectrum of fluctuations on the long string.  Here we rederive the results of Ref.~\cite{Siemens:2001dx,Siemens:2002dj} in perhaps a simpler way.  Our result for the exponent of $G\mu$ differs somewhat from Ref.~\cite{Siemens:2002dj}.   This difference arises in going from the periodic wave train calculation of Ref.~\cite{Siemens:2001dx} to the random distribution on the actual network.  Also, we use input from our model~\cite{Polchinski:2006ee} and from simulations~\cite{Martins:2005es} to determine the actual power spectrum on the long string.

In Sec.~3 we discuss loop production.  We use the results from~Sec.~2 to find the cutoff on loop sizes.  It is notable that in our model these smallest loops will be highly relativistic.  We then consider the recent simulations of Ref.~\cite{Vanchurin:2005pa,Olum:2006ix}, which are able to cover the longest time scales, and which show large loops.  These simulations also show a population of small loops near the cutoff scale, which are claimed to be transient.  We instead argue on several grounds that these smaller loops are a real effect, so that the true loop distribution has {\it two} peaks.  One scales at $\alpha \sim 0.1$, while the second is presumably at the scale set by gravitational radiation.  We discuss how the resulting picture affects current limits and future sensitivity of searches for cosmic strings.  In particular, Refs.~\cite{Olum:2006at,Hogan:2006we,Siemens:2006yp} showed that the $\alpha \sim 0.1$ loops produce a much larger stochastic background than in previous network scenarios, tightening existing bounds and increasing the potential for discovery.  If these loops represent only a fraction of the total string length then this effect is somewhat reduced, but still significant.  We also discuss the gravitational wave signatures of the small loops, and the interaction between the short distance structure and the string cusps.

\sect{Gravitational radiation}

\subsection{Model for small scale structure}

Let us first review some of the results of Ref.~\cite{Polchinski:2006ee} for the small scale structure on a long cosmic string.
The motion of a cosmic string in a three-dimensional target space can be described by a map ${\bf x}(\sigma,t)$, where $\sigma$ parametrizes the distance along the string and $t$ is the FRW time.  Denote $\partial_\sigma$ by a prime and $a(t) \partial_t = \partial_\tau$ by a dot (where $a(t)$ is the scale factor and  $\tau$  is the conformal time), and define $\epsilon \equiv \sqrt{ {\bf x}'^2 / (1-\dot{\bf x}^2) }$.  Then in transverse gauge, $\dot{\bf x} \cdot {\bf x}'=0,$ the quantities ${\bf p}_\pm = \dot{\bf x} \pm \frac{1}{\epsilon} {\bf x}'$ are unit vectors.   In flat space these describe simply left- and right-moving waves.  In an expanding universe the equations of motion derived from the Nambu action introduce a correction to the wave equation which is proportional to $\dot a / a$.

To better understand the small scale behavior of cosmic string networks a model was developed in~\cite{Polchinski:2006ee} that focused on the two-point functions $\left< {\bf p}_\pm(\sigma, t) \cdot {\bf p}_\pm (\sigma', t) \right>$.  This correlator must vanish at scales larger than the horizon size due to causality, while it encodes the short distance structure observed at smaller scales.  Assuming that stretching of the network as spacetime expands gives the dominant behavior at scales small compared to the horizon but long compared to the gravitational radiation scale, the two-point function was determined to be
\begin{equation}
\langle {\bf p}_\pm (\sigma,t) \cdot {\bf p}_\pm (\sigma',t) \rangle
\approx 1 - {\cal A} (l/t)^{2\chi}
\ ,
\label{pp}
\end{equation}
where $ l = a(t) \epsilon(t) \left| \sigma - \sigma' \right| $ is the physical separation (weighted by the local value of $\gamma$ along the string).  This quantity is a function of the ratio $l/t$ only and so it scales properly.  The exponent $\chi$ is a function of the rate of change of the scale factor and also the RMS velocity of the strings, $\overline{v}^2$.  For a power law expansion of the form $a(t) \propto t^\nu$ one finds~\cite{Polchinski:2006ee} 
\begin{equation}
\chi = \frac{\nu \overline{\alpha}}{1 - \nu \overline{\alpha}} \ ,\quad
\overline{\alpha} \equiv 1 - 2 \overline{v}^2
\ .
\label{chi}
\end{equation}
In a flat spacetime the virial theorem imposes $\overline{v}^2 = 1/2$ (in natural units) but due to expansion $\overline{v}^2 < 1/2$.  Thus, for a decelerating universe the exponent $\chi$ is a positive number, taking the specific values $\chi\sim 0.10$ in the radiation dominated era and $\chi\sim 0.25$ in the matter dominated era.

These exponents are in reasonable agreement with the two-point functions given in Ref.~\cite{Martins:2005es} over a range of scales below the horizon size, with the normalization $ {\cal A}$ found to be $\sim 0.60$ in both the matter and radiation epochs.  At smaller scales the simulations seem to indicate a larger exponent $\chi\sim 0.5$, corresponding to the functions ${\bf p}_\pm$ mapping out a random walk on the unit sphere.  We are unable to identify a physical mechanism that would produce this effect, and believe that the discrepancy is due to transients in the simulations.  In any event, our discussion below just uses the general power law form~(\ref{pp}), and one can insert any assumed values for $\chi$ and $ {\cal A}$.

Also useful for the computations in the next section is the small fluctuation approximation introduced in~\cite{Polchinski:2006ee}.  In this framework the structure on a small segment is expressed as a large term that is constant along the segment plus a small fluctuation:
\begin{equation}
{\bf p}_+ (\sigma,t) = {\bf P}_+ (t) + {\bf w}_+ (\sigma,t) - \frac{1}{2} {\bf P}_+ (t)
w_+^2  (\sigma,t) + \ldots\ ,  \label{smallf}
\end{equation}
with $P_+(t)^2 = 1$ and ${\bf P}_+ (t) \cdot {\bf w_+} (\sigma,t) = 0$.  An equivalent equation can also be written for the right-moving vector ${\bf p}_-$.  Equation~(\ref{pp}) can then be expressed in terms of the fluctuations as
\begin{equation}
1 - \langle {\bf p}_\pm (\sigma,t) \cdot {\bf p}_\pm (\sigma',t) \rangle
= \frac{1}{2} \langle \left[ {\bf w}_\pm (\sigma,t) - {\bf w}_\pm (\sigma',t) \right]^2 \rangle  + O(w^4_\pm)
\approx  {\cal A} (l/t)^{2\chi}
\ .
\label{ww}
\end{equation}

\subsection{Radiation and back-reaction}

Gravitational radiation is important on scales short compared to the Hubble time and so we can use the flat metric diag$(-1,+1,+1,+1)$, and also make the gauge choice $a\epsilon = 1$.\footnote{This choice is possible when we consider scales small compared to the Hubble time.  In Sec.~2.4, when we again consider cosmological evolution, we must reintroduce $a(t)$ and $\epsilon(t)$.}  The modes are then functions only of $u = t + \sigma$ or $v = t - \sigma$, so that
\begin{equation}
\langle {\bf p}_+ (u) \cdot {\bf p}_+ (u') \rangle
\approx 1 -  {\cal A} [(u-u')/t]^{2\chi} \ , \quad
\langle {\bf p}_- (v) \cdot {\bf p}_- (v') \rangle
\approx 1 -  {\cal A} [(v-v')/t]^{2\chi} \ ;
\end{equation}
(the explicit $t$ on the right-hand side is effectively a fixed parameter, varying only over the Hubble time).

The energy radiated per unit solid angle in the $\bf k$ direction is given by~\cite{Sakellariadou:1990ne,Hindmarsh:1990xi,Siemens:2001dx}
\begin{equation}
\frac{d\Delta E}{d\Omega} = \frac{G\mu^2}{16\pi^2} \int_0^{\infty} d\omega \, \omega^2
\left\{ |A|^2 |B|^2 + |A^* \cdot B|^2 - |A \cdot B|^2 \right\}
\ ,
\label{energy}
\end{equation}
with the left- and right-moving contributions given respectively by
\begin{eqnarray}
A^\mu(k) &=& \int_{-\infty}^{+\infty} du \, p_+^\mu(u) 
\exp\left\{ { -\frac{i}{2} \int_0^u du' \, k \cdot p_+(u') } \right\} \ , \nonumber\\
B^\mu(k) &=& \int_{-\infty}^{+\infty} dv \, p_-^\mu(v) 
\exp\left\{  -\frac{i}{2} \int_0^v dv' \, k \cdot p_-(v')  \right\}
\ .
\label{AB}
\end{eqnarray}
Here, $k^\mu$ is the gravitational radiation wave vector, and $p^\mu_\pm$ are null 4-vectors whose time component is identically 1 and with ${\bf p}_\pm$ being the unit vectors above.  These are related to the standard notation 
\begin{equation}
x^\mu(u,v) = \frac{1}{2} [ a^\mu(u) + b^\nu(v) ]
\end{equation}
 via $p^\mu_+(u) = a^{\mu\prime}(u)$ and $ p^\mu_-(v) = b^{\mu\prime}(v)$.
In the spirit of the small fluctuation approximation we orient a given segment of string mainly along the $z$-axis and consider fluctuations ${\bf w}_\pm$ in the perpendicular $xy$-plane.  Thus,
\begin{equation}
p^\mu_+(u) \simeq \left( 1, {\bf w}_+(u), \sqrt{1-w_+^2(u)} \right) \ , \quad
p^\mu_-(v) \simeq \left( 1, {\bf w}_-(v), - \sqrt{1-w_-^2(v)} \right)
\ .
\label{pw}
\end{equation}
We can ignore the last two terms in equation~(\ref{energy}): they are equal when $B^\mu$ is real, and so cancel when ensemble-averaged over the short distance structure.

Let us review the argument of Refs.~\cite{Siemens:2001dx,Siemens:2002dj}.  First linearize the modes~(\ref{AB}) in the oscillations, so the exponential factors become $e^{i k_a u}$ and 
$e^{i k_b v}$ respectively, with 
\begin{equation}
\omega = k_a + k_b \ , \quad k_z = k_b - k_a
\ .
\label{conservation}
\end{equation}
Here $\omega$ and $k_z$ are the frequency and wavenumber of the radiation.
Then the ensemble averages in the linearized approximation ($\approx$) are
\begin{eqnarray}
\langle |A(k)|^2 \rangle &\approx& l_u \int_{-\infty}^{+\infty} du \, e^{i k_a u
- \varepsilon |u|}
\left( \langle {\bf p}_+ (u) \cdot {\bf p}_+ (0) \rangle - 1 \right) =  l_u {\cal A} \, c_\chi \,  k_a^{-(1+2\chi)} t^{-2\chi} \ ,\nonumber\\
\langle |B(k)|^2 \rangle &\approx&  l_v {\cal A} \, c_\chi \, k_b^{-(1+2\chi)} t^{-2\chi}  \ .
\end{eqnarray}
Here $c_\chi =  2 \sin (\pi\chi) \, \Gamma(1+2\chi)$, taking the values 1.25 and 0.57 in the matter- and radiation-dominated eras respectively.  The convergence factor $e^{-\varepsilon |u|}$ accounts for the decay of the correlations on horizon scales; its detailed form is unimportant at shorter scales.  The factors $l_u$ and $l_v$ are volume regulators: we artificially cut the oscillations off to produce finite trains, but these regulators of course drop out when we consider rates per unit time and length.  Using 
$
d\omega \, d\cos\theta = 2 {dk_a dk_b}/({k_a + k_b}) \ ,
$
we can write the total energy radiated as
\begin{eqnarray}
\langle \Delta E \rangle &=& \frac{G\mu^2}{4 \pi} \int_0^\infty\!\!  dk_a \int_0^\infty\!\! dk_b \, (k_a + k_b)
\langle |A(k_a)|^2 \rangle  \langle |B(k_b)|^2 \rangle \nonumber\\
&\approx&
 \frac{G\mu^2  {\cal A}^2 c_\chi^2}{ 4\pi } l_u l_v \int_0^\infty\!\! dk_a \int_0^\infty\!\! dk_b \, \frac{k_a + k_b}
{ (k_a k_b)^{1+2\chi} t^{ 4\chi} } 
\ .
\label{power2}
\end{eqnarray}
The volume of the world-sheet is $V = \frac{1}{2}  l_u l_v$, so this translates into a power per unit length
\begin{equation}
\left\langle \frac{dP}{dz} \right\rangle \approx
 \frac{G\mu^2  {\cal A}^2 c_\chi^2}{ 2\pi } \int_0^\infty\!\! dk_a \int_0^\infty\!\! dk_b \, \frac{k_a + k_b}
{ (k_a k_b)^{1+2\chi} t^{ 4\chi} } 
\ .
\label{power3}
\end{equation}
The total energy of the wavetrains is
\begin{equation}
E \approx {\mu} \int_0^\infty \frac{dk_a}{2\pi}\,|A(k_a)|^2  +{\mu} \int_0^\infty \frac{dk_b}{2\pi}\,|B (k_b)|^2 \ .
\end{equation}
Isolating a momentum range $dk_a$, we have
\begin{equation}
\Delta  |A(k_a)|^2 \approx - \frac{G\mu k_a}{2}  |A(k_a)|^2 \int_0^{\infty} dk_b \, 
\langle |B(k_b)|^2 \rangle \ . \label{Adecay}
\end{equation}
We have used the fact that momentum conservation determines that the energy coming from the $A$ and $B$ modes is in the proportion $k_a$ to $k_b$.  A given point $u$ interacts with the $v$ train for a time $\frac{1}{2} \Delta v$, so the rate of decay becomes
\begin{equation}
\frac{d}{dt}  |A(k_a)|^2 \approx
-{G\mu}  k_a  |A(k_a)|^2 {  {\cal A} c_\chi} \int_0^\infty
 \frac{ dk_b} {k_b^{1+2\chi} t^{ 2\chi} } 
\ . \label{decay1}
\end{equation}
The integral is dominated by the lower limit: most of the energy loss from the left-moving mode $k_a$ comes from its interaction with right-moving modes of much longer wavelength.
Cutting off the lower end of the $k_b$ integral at the horizon scale, $k_b \sim 1/t$, we find that the decay rate is of order $G\mu  k_a$.  This is faster than the Hubble time for 
\begin{equation}
k_a > O(G\mu t)^{-1}\ , \label{naive}
\end{equation}
meaning that modes with wavelengths $\lambda < O(G\mu t)$ are exponentially suppressed.  Thus, Eq.~(\ref{naive}) reproduces the usual (naive) estimate for the gravitational damping length.

\subsection{Radiation and back-reaction improved}

The calculation above implies that the energy loss from the left-moving modes $k_a$ comes primarily from their interaction with much longer right-moving modes at the horizon scale.  Ref.~\cite{Siemens:2001dx} argued that this was paradoxical, because a wave that encounters an oncoming perturbation with much longer wavelength is essentially traveling on a straight string, and such a wave will not radiate.
They showed that this paradox arose due to the neglect of the exponential terms in the modes~(\ref{AB}).  With their inclusion, the radiation is exponentially suppressed when the ratio of $k_a$ to $k_b$ becomes sufficiently large.

Ref.~\cite{Siemens:2001dx} considered monochromatic wavetrains 
\begin{equation}
w_+^x(u) = \epsilon_a \cos (k_a u)\ , \quad w_-^x(v) = \epsilon_b \cos (k_b v)\ ,
\end{equation}
and found that the gravitational radiation was suppressed unless
\begin{equation}
k_b/ \epsilon_b^2 \gsim  k_a\ ,\quad k_b \lsim k_a/\epsilon_a^2\ . \label{range}
\end{equation}
The first of these relations will cut off the integral~(\ref{decay1}) below the horizon scale, but first
we need to extend it to the incoherent spectrum on a long cosmic string.  We will do this in a systematic way below, but we can anticipate the answer.  For a continuous spectrum, of course, a single frequency makes a contribution of measure zero.  The natural extrapolation is then to replace $\epsilon_{a}, \epsilon_b$ with the RMS average in bands of frequencies $k \sim k_a, k_b$.  That is,
\begin{eqnarray}
\epsilon_b^2 &\sim& \int_{k_b/\Lambda}^{\Lambda k_b} \frac{dk}{2\pi} \int_{k_b/\Lambda}^{\Lambda k_b} \frac{dk'}{2\pi} \, \tilde{\bf w}_-(k) \cdot  \tilde{\bf w}_-(k')^*
\nonumber\\
&=& -\frac{1}{2} \int_{-\infty}^\infty dv \int_{-\infty}^\infty dv' \, g(v) g(v')^*  [{\bf w}_-(v) - {\bf w}_-(v')]^2 \nonumber\\
&\sim & (k_b t)^{-2\chi}\ ;
\end{eqnarray}
here $\Lambda = \sqrt{e}$ is a plausible choice, corresponding to integrating over one unit of log frequency, but the precise value does not matter.  In the second line the filter function 
\begin{equation}
g(v) = \int_{k_b/\Lambda}^{\Lambda k_b} \frac{dk}{2\pi} e^{-i k v}
\end{equation}
has a width of order $1/k_b$ and a magnitude of order $k_b$.  In the third line, we have used the two-point function~(\ref{ww}).
Thus,
\begin{equation}
(k_b t)^{1 + 2\chi} \gsim k_a t\ . \label{socut}
\end{equation}

Using this lower cutoff in the decay rate~(\ref{decay1}) gives
\begin{equation}
\frac{d}{dt}  |A(k_a)|^2 \sim
- {G\mu} (k_a t)^{1/(1+2\chi)} t^{-1}  |A(k_a)|^2 \equiv - \frac{|A(k_a)|^2}{\tau_{\rm GR}(k_a)} \ .
\end{equation}
This is faster than the Hubble rate for
\begin{equation}
k_a \gsim (G\mu)^{-(1+2\chi)} t^{-1}\ . \label{true}
\end{equation}
The gravitational length scale is reduced from the naive $O(G\mu t)$ by a factor $(G\mu)^{2\chi}$. The qualitative conclusion is the same as in Ref.~\cite{Siemens:2002dj}, but the exponents do not seem to agree.  The result there was $k_a > (G\mu)^{-(1 + 2\beta)/2} t^{-1}$.  It is not clear whether $\beta$ is to be identified with $\chi+ \frac{1}{2}$ (as suggested by Eq.~(21) of Ref.~\cite{Siemens:2002dj}, compared with Eq.~(\ref {decay1}) above) or with $\chi$ (as suggested by \cite{Siemens:2002dj} Eq.~(31)), but in either case the exponents differ.  This stems from a different method of converting from the single-mode result to the continuous spectrum; our exponent will be borne out by the more formal treatment below.  

Finally, recalling Sec.~2.1, our analytic model~\cite{Polchinski:2006ee} gives for the exponent $1+2\chi$ the values 1.2 in the radiation era and 1.5 in the matter era;  using instead the simulations~\cite{Martins:2005es} would yield an exponent around 2.0 at the shortest scales in both eras.\footnote{Numerically, there are two changes from Ref.~\cite{Siemens:2002dj}: the corrected expression for the exponent, and a more accurate estimate of the power spectrum.  Ref.~\cite{Siemens:2002dj} effectively estimated the latter using $\bar v^2 = 0$, so that their Eq.~(31) is $\beta = \chi|_{\bar v^2 = 0} = \nu/(1-\nu).$}

The necessary correction to the naive radiation formula comes from the previously neglected exponential factor in $B^\mu(k)$, Eq.~(\ref{AB})~\cite{Siemens:2001dx}.  Expanding the exponential to second order in the fluctuations gives
\begin{eqnarray}
|B(k) |^2 &=& \int_{-\infty}^{+\infty} dv \int_{-\infty}^{+\infty} dv' \,( {\bf p}_-(v) \cdot  {\bf p}_-(v') - 1 )
\nonumber\\
&&\quad
\exp\left\{  i k_b (v'-v)  - \frac{i {\bf k}_\perp}{2} \cdot \int_v^{v'} dv'' \,  {\bf w}_-(v'') 
- \frac{i k_z}{4} \int_v^{v'} dv''\, w_-^2(v'')  \right\}
\ .\quad
\label{Bexp}
\end{eqnarray}
Noting that $k_\perp^2 = 4k_a k_b$, when the first term in the exponent is of order one, the second and third terms in the exponent are respectively of order $(k_a t)^{1/2} (k_b t)^{-1/2- \chi}$ and $k_a t (k_b t)^{-1-2\chi}$.  Thus, as $k_b$ decreases they become important at precisely the scale~(\ref{socut}) where the small-amplitude calculation is expected to break down.  The linearized form for $A(k)$ remains valid, so we can simply insert the corrected form~(\ref{Bexp}) into the decay rate~(\ref{Adecay}).

By completing the square, the exponent in Eq.~(\ref{Bexp}) can be written as
\begin{eqnarray}
&&i \frac{(v'-v)}{4 {k_a} }  \left( {{\bf k}_\perp} -  \frac{k_a}{\sqrt{ v'-v } } \int_v^{v'} dv'' \, {\bf w}_-(v'') \right)^2  \nonumber \\
&& \qquad \qquad - \frac{i k_z}{8 (v' - v)}  \int_v^{v'} dv'' \int_v^{v'} dv'''\, [{\bf w_-}(v'') - {\bf w_-}(v''')]^2  \ .
\end{eqnarray}
Converting $\int dk_b \to (4 k_a)^{-1} \int d k_\perp^2 \to 
(4\pi k_a)^{-1} \int d^2 k_\perp$, we have at fixed $k_a$ the integral
\begin{eqnarray}
\int_0^\infty dk_b \, |B(k) |^2 &=& \int_{-\infty}^{+\infty} dv \int_{-\infty}^{+\infty} dv' \,
\frac{i ( {\bf p}_-(v) \cdot  {\bf p}_-(v') - 1 )}{v'-v}
\nonumber\\
&&\quad
\exp\left\{ 
- \frac{i k_z}{8 (v' - v)} \int_v^{v'} dv'' \int_v^{v'} dv'''\, [{\bf w_-}(v'') - {\bf w_-}(v''')]^2  \right\}
\  .\quad
\label{Bexp2}
\end{eqnarray}
We have used $-k_z \approx k_a \gg k_b$, since this is the regime where the correction is important.

We now approximate, replacing both the exponent and the prefactor with their mean values from Sec.~2.1.  Then
\begin{eqnarray}
\int_0^\infty dk_b \, \langle |B(k)|^2 \rangle &=& 2 {\cal A} l_v\,  {\rm Im} \int_{-\infty}^{+\infty} dv \,
\frac{v^{2\chi-1}}{ t^{2\chi}}
\exp\left\{ 
\frac{i k_a {\cal A}}{4 v} \int_0^{v} dv'' \int_0^{v} dv'''\,\frac{ |v''-v'''|^{2\chi}}{t^{2\chi}} \right\}
\nonumber\\
&=&  2 l_v  b_\chi  ( k_a t )^{-2\chi/(1+2\chi)}  \ .
\label{Bexp3}
\end{eqnarray}
Here
\begin{equation}
b_\chi =  2 \sin\left( \frac{\pi\chi}{1+2\chi} \right) \Gamma\left( \frac{ 2\chi }{1 + 2\chi} \right) [4 (1+\chi)]^{2\chi/(1 + 2\chi)} \left[ \frac{\cal A} {1+2\chi}\right]^{1/(1 + 2\chi)} 
\end{equation}
is 2.5 in the matter era and 2.1 in the radiation era.  We can improve this approximation if we make the assumption that the ensemble is gaussian.  The first order correction, keeping contractions between the prefactor and the exponent, involves a straightforward calculation; its inclusion simply renormalizes the constant $b_\chi \to (1+ \epsilon_\chi)b_\chi$, with
\begin{equation}
\epsilon_\chi = \chi (1+\chi) \left[ \frac{\Gamma(1+2\chi)^2}{\Gamma(2+4\chi)} - \frac{1}{2(1+4\chi)} \right] \ .
\end{equation}
However, this represents a small correction since $\epsilon_\chi = 0.035$ in the radiation epoch and $\epsilon_\chi = 0.045$ in the matter epoch.
Thus, to good approximation,
\begin{equation}
\frac{1}{\tau_{\rm GR}(k_a)} = b_\chi G\mu ( k_a t )^{1/(1+2\chi)} t^{-1} \ .
\label{tauGR}
\end{equation}
This is faster than the Hubble rate for the scales~(\ref{true}), as deduced earlier.

\subsection{The long-string two-point function}

We can now improve our earlier result for the two-point function~\cite{Polchinski:2006ee} through the inclusion of gravitational radiation.  Define
\begin{equation}
H(\kappa,t) = a(t) \epsilon(t) \int_{-\infty}^{\infty} d\sigma\, e^{-i \kappa\sigma}  \langle {\bf w}_+(\sigma, t)
\cdot {\bf w}_+(0, t) \rangle\ ,
\label{Hkappa}
\end{equation}
where $\sigma$ is related to the actual length along the string by $l = a(t) \epsilon(t) \sigma$; here  $a = t^\nu$  and $\epsilon(t) = a^{-2\bar v^2}$.   Combining the stretching given by Eq.~2.26 of Ref.~\cite{Polchinski:2006ee} with the gravitational radiation found above gives
\begin{equation}
\frac{d}{dt} H(\kappa,t) = - \frac{ \dot a}{a} \bar\alpha H(\kappa,t) - \frac{1}{\tau_{\rm GR}(\kappa/a \epsilon)} H(\kappa,t)\ .
\end{equation}
This is readily integrated, to give
\begin{equation}
\ln H(\kappa,t) = - \bar\alpha \ln a(t) + \ln f(\kappa) - b_\chi (1+\chi)(1+2\chi) G\mu \kappa^{1/(2\chi+1)} t^{1/(1+\chi)(1+2\chi) } \ .
\end{equation}
The function $f(\kappa)$ is obtained by matching onto the known result~(\ref{ww}) at early times, giving $f(\kappa) = {\cal A} c_\chi \kappa^{-1-2\chi}$.  Expressing the result in terms of the physical momentum scale $k = \kappa /a(t)\epsilon(t)$ gives
\begin{equation}
J(k,t) = H(\kappa,t) =  {\cal A} c_\chi k^{-1} (kt)^{-2\chi} \exp\left[ -b_\chi (1+\chi)(1+2\chi) G\mu (kt)^{1/(1+2\chi)} \right] \ .
\label{2pf}
\end{equation}

\begin{figure}
\center \includegraphics[width=30pc]{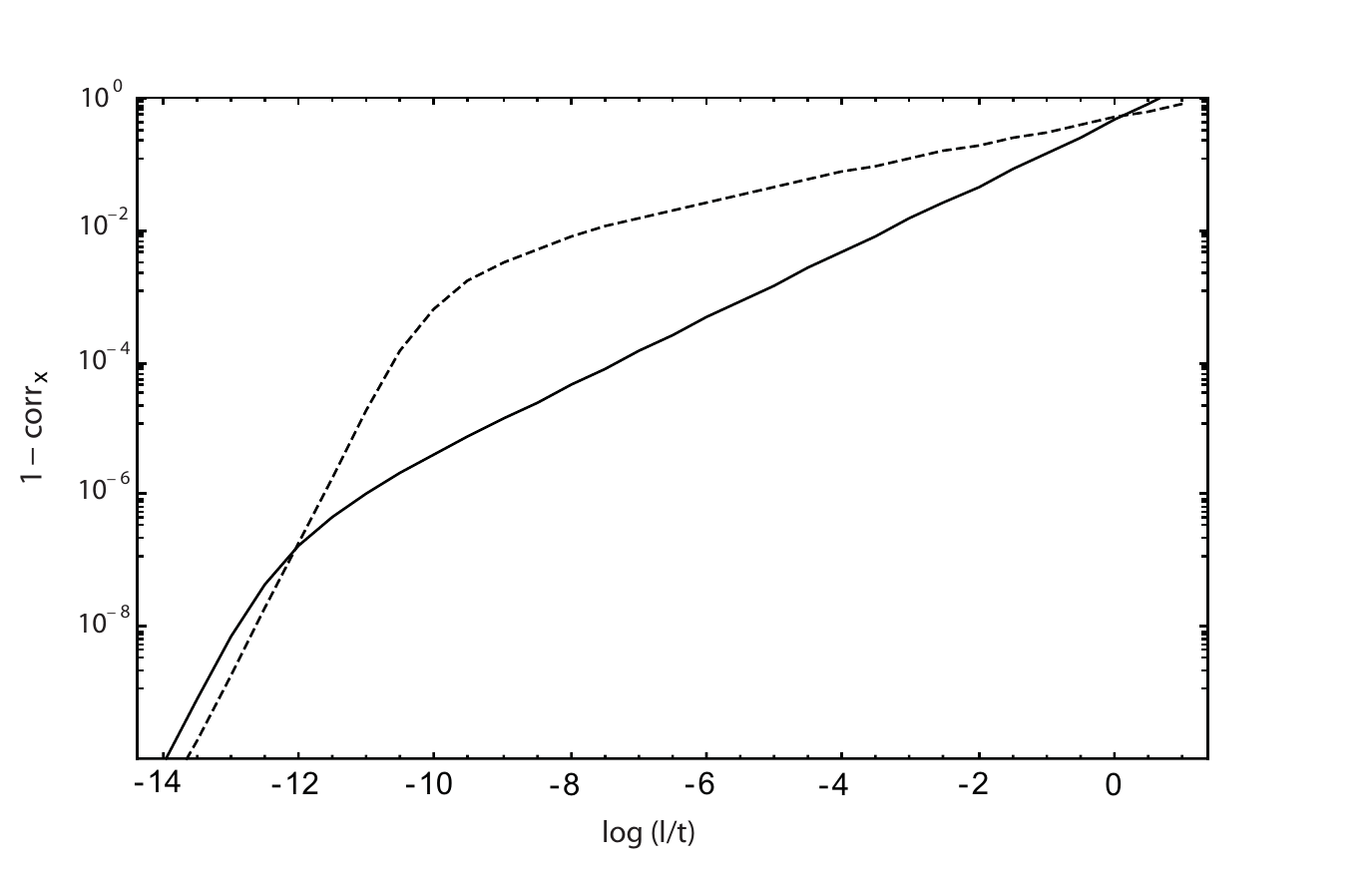}
\caption[]{ The effect of the inclusion of gravitational radiation on the long-string two-point function is to smooth the string on scales below $\sim 20(G \mu)^{1+2\chi}t$.  The solid line corresponds to the matter epoch and the dotted line to the radiation epoch.  The value $G \mu = 10^{-9}$ has been adopted for illustrative purposes.} \label{2pf plot}
\end{figure}

Thus, unless we consider modes with such short wavelength that the exponential suppression factor kicks in, we have $J(k) \sim k^{-(1+2\chi)}$.  At length scales smaller than $\sim (G \mu)^{1+2\chi} t$ this behavior is altered as a result of the inclusion of gravitational radiation, in such a way to make the two-point function smoother.   In Fig.~\ref{2pf plot} we show the numerical Fourier transform of $J(k,t)$, in terms of the correlation~\cite{Martins:2005es, Polchinski:2006ee}
\begin{equation}
1 - {\rm corr}_x  \approx  \frac{1 - \langle {\bf p}_+(l,t) \cdot {\bf p}_+(0,t) \rangle}{2(1-\overline{v}^2)}  =  \frac{ \langle w_+(l,t)^2 \rangle - \langle {\bf w}_+(l,t) \cdot {\bf w}_+(0,t) \rangle}{2(1-\overline{v}^2)} \ .
\end{equation}

The smoothing at short distance seen in Fig.~\ref{2pf plot} resembles that found in the numerical simulations of Ref.~\cite{Martins:2005es}, but here it is due to an additional physical effect, gravitational radiation, whereas in Ref.~\cite{Martins:2005es} such smoothing was found even without gravitational radiation.\footnote{As an aside, the correlator in Fig.~\ref{2pf plot} goes as $l^2$ at short distance, corresponding to an ensemble of smooth functions, whereas it appears to be close to $l^1$ in Ref.~\cite{Martins:2005es}, corresponding to a random walk in $ {\bf p}_\pm$.}  Again, it remains to be seen whether the result of Ref.~\cite{Martins:2005es} is a real effect or a transient.  In Fig.~\ref{2pf plot} we have assumed the latter, so that the two-point function follows a simple power law down to the gravitational radiation scale.

\sect{Loops}

\subsection{Small loops}

The exponential falloff in the world-sheet two-point function at large wavenumber implies that the strings are very smooth at short distance.  Inspection of Fig.~\ref{2pf plot} indicates that this smoothing sets in at
\begin{equation}
l_{\rm c} \approx 20  (G \mu)^{1+2\chi} t \label{lsize}
\end{equation}
in both the matter and radiation epochs.  The exponent we obtain is smaller than the estimates in~\cite{Siemens:2002dj}: $1.2$ in the radiation dominated era and $1.5$ in the matter dominated era here, compared to $3/2$ and $5/2$ respectively.  The difference is primarily from using a more accurate model of the effect of stretching, leading to a power spectrum with a slower falloff.  The length~$l_{\rm c}$ is the approximate size of the smallest loops.  In particular, the divergent loop production found in Ref.~\cite{Polchinski:2006ee} is cut off at this scale.

It is customary to parameterize the size of small loops as $l = \epsilon \Gamma G\mu t$ with $\Gamma \sim 50$.  The parameter $\epsilon$ corresponds to the lifetime of the loop in units of the FRW time, if gravitational radiation is the principal decay mode.  Thus we have found 
\begin{equation}
\epsilon \approx 0.4 (G\mu)^{0.5}
\end{equation}
in the matter era, while for the radiation era
\begin{equation}
\epsilon \approx 0.4 (G\mu)^{0.2} \ .
\end{equation}

However, for the phenomenology of the small loops there is another important property, first noted in Ref.~\cite{Polchinski:2006ee}: they move {\it extremely} fast.  The point is that the functions ${\bf p}_+(u)$ and ${\bf p}_-(v)$ must have equal mean values on the loop (the condition for the loop to form).  If the loop is small, so these functions have limited range, this implies that both remain near one point on the sphere, and the velocity (the average of ${\bf p}_+(u)$ and ${\bf p}_-(v)$ over the loop) will be very close to unity.  From Eq.(4.34) of Ref.~\cite{Polchinski:2006ee}, the Lorentz contraction factor is
\begin{equation}
\gamma = (l/t)^{-\chi} [( 1+\chi)(1+ 2\chi) / {\cal A} ]^{1/2} \ .
\end{equation}
This is $\approx 0.8 (G\mu)^{-0.37}$ in the matter era, and is of order $10^3$ for tensions of interest;  in the radiation era it is $\approx 1.0 (G\mu)^{-0.12}$ and is of order $10^1$.
Since this factor is so large, we must be careful to recall that the specific definition of length is 
\begin{equation}
dl = a(t) \epsilon(t) d\sigma =  a(t) \sqrt{ {\bf x}'^2 / (1-\dot{\bf x}^2) }  d\sigma
\end{equation}
and so $l$ is $\gamma$ times the rest frame length $l_{\rm rest}$.  It is equal to twice the inverse period of the loop in the FRW rest frame.

To see one effect of the boost, in Ref.~\cite{Siemens:2006yp} the total stochastic gravitational wave spectrum from small loops was considered.  The frequency spectrum simply scales inversely as the size of the loops if they are at rest, $\omega \propto 1/l_{\rm rest}$.  In the FRW rest coordinate this frequency is reduced by a factor of $\gamma$.  However, the large boost pushes essentially all of the radiation into the forward direction, where the frequency is enhanced by a factor $(1-v)^{-1} \approx 2 \gamma^2$.  Thus the gravitational radiation spectrum corresponds to an effective loop size
\begin{equation}
\epsilon_{\rm eff} \approx \gamma^{-1} \epsilon_{\rm rest}/2  \approx \gamma^{-2} \epsilon/2\ .\label{effeps}
\end{equation}
This is $\epsilon_{\rm eff} \approx 0.3 (G\mu)^{1.25}$ in the matter era, meaning that $l_{\rm eff} \approx 15 (G\mu)^{2.25}$.  In the radiation era $\epsilon_{\rm eff} \approx 0.2 (G\mu)^{0.44}$, meaning that $l_{\rm eff} \approx 10 (G\mu)^{1.44}$.  We should note that Ref.~\cite{Siemens:2006yp} considered the sensitivity in the whole $(G\mu , \epsilon)$ plane, and we are simply highlighting one line in this plane.  It is notable that Advanced LIGO is more sensitive than LISA to these smallest loops, because their small size puts them above the LISA frequency range.  In fact, for interesting values of the tension the loops will produce a periodic signal in the LIGO frequency band, which may be observable at Advanced LIGO~\cite{wip}.

\subsection{Loops: large and small}

Finally, let us discuss how these small loops might fit into the overall picture of the string network.  In Ref.~\cite{Polchinski:2006ee} we found that loop production from long strings diverges (as measured by total string length) at small sizes.  To match the scaling value for the rate of long string loss to loops we would need to cut this divergence off at $l \sim 0.18 t$; this is small in the sense of being an order of magnitude below the horizon length, but very large compared to the scales that we have discussed thus far.  However, the emission of loops from long strings does not end the story, since the divergent loop production continues on the large loops themselves.  We were unable to go further analytically, but conjectured that loops would fragment extensively until they reached a non-self-intersecting configuration at a scale we guessed to be a few orders of magnitude smaller, $l \sim 10^{-3}t$ to $10^{-4}t$.  This appeared to be consistent with the simulations of Refs.~\cite{Ringeval:2005kr,Martins:2005es}.  If this were the case then the gravitational radiation smoothing that we have discussed would be largely irrelevant, because loop production would already be cut off at much larger scales.

However, Refs.~\cite{Vanchurin:2005pa,Olum:2006ix} have recently been able to simulate over longer time scales, and argue for a rather different picture.  Namely, they find two populations of loops, one of which scales at $\alpha \equiv l/t \sim 0.1 $, and a second which remains at the simulation cutoff length.  They argue that $\alpha \sim 0.1$ is the actual loop size, and that the small loops are a transient effect in the simulations and will not be present in the real networks.

We would argue that this interpretation of the small loops as transients is implausible.  First, there is no sign in Refs.~\cite{Vanchurin:2005pa,Olum:2006ix} that the total fraction of string length going into the small loops is diminishing in time.  Second, if this were a transient effect then its natural time scale should be small in proportion to the size of the loops.  Instead, these loops are still being produced after several Hubble times, showing that some longer-distance process is continuing to feed the small loop production.  Indeed, we have identified this mechanism already~\cite{Polchinski:2006ee}: the small scale structure on long strings and large loops, which originates at the horizon length and then is carried down to smaller relative lengths by the expansion of the universe, leads to production of loops on arbitrarily small scales. 

Thus we are led to suggest the following picture.  After a large loop is formed, production of smaller loops continues near its cusp regions.  In the end, there is a large loop without cusps or self-intersections, reduced in size by the excision of the cusp regions.  There is also a population of much smaller loops; based on the non-scaling seen thus far~\cite{Vanchurin:2005pa,Olum:2006ix}, we conjecture that these will be at the gravitational radiation length.\footnote{In the absence of gravitational wave smoothing, these could even be at the scale set by the thickness of the string~\cite{Bevis:2006mj}.}  Inspection of Fig.~4 of Ref.~\cite{Olum:2006ix} suggests that as much as 80\% of the total string length goes into the small loops, with 20\% remaining in the large loop.

As has been discussed in Refs.~\cite{Olum:2006at,Hogan:2006we,Siemens:2006yp},  the $\alpha \sim 0.1$ loops produce a much larger stochastic background than in previous network scenarios, tightening existing bounds and increasing the potential for discovery.  If these loops represent only a fraction of the total string length then this effect is reduced.  For loops produced and decaying during the radiation era, the total energy density today is proportional to $(G\mu\alpha)^{1/2}$.
Thus, if only 20\% of the string length goes into large loops, the bound on $G\mu$ is weaker by a factor of 25.  For example, the pulsar timing bound $G\mu < 8 \times 10^{-9}$ for reconnection probability $p=1$, shown in Fig.~2 of Ref.~\cite{Siemens:2006yp}, would be weakened to $G\mu < 2 \times 10^{-7}$, very close to the current CMB bound~\cite{Pogosian:2006hg}.  Foreseeable improvements in pulsar timing, e.g.~\cite{Jenet:2006sv}, will improve the sensitivity by several orders of magnitude in $G\mu$.

\subsection{Cusps}

Our work also allows us to address an old question, the interplay between the small scale structure and the cusps~\cite{Siemens:2001dx}.  A cusp arises when the functions ${\bf p}_+(u)$ and ${\bf p}_-(v)$ cross on the unit sphere~\cite{Albrecht:1989mk}.  The {\it size} of the cusp is the inverse of the `velocity' of the functions ${\bf p}_{\pm}$ when they cross, their rate of change with respect to the length along the string.  Now, the average velocity of these functions on the long strings is (going again to the gauge $\epsilon(t) = a(t) = 1$)
\begin{equation}
V^2 = \langle  \partial_\sigma {\bf p}_+(\sigma,t) \cdot  \partial_\sigma {\bf p}_+(\sigma,t) \rangle
= \int \frac{d\kappa} {2\pi} \kappa^2 H(\kappa,t) = d_\chi(G\mu)^{-2(1-\chi)(1+2\chi)}  t^{-2} \ , \label{speed}
\end{equation}
where
\begin{equation}
d_\chi = {\cal A} c_\chi \frac{1+2\chi}{2\pi} \Gamma[2(1-\chi)(1+2\chi)]\, [ b_\chi (1+\chi)(1+2\chi) ]^{2(\chi - 1)(1+2\chi)} \label{V2}
\end{equation}
takes the value 0.006 in the matter era and 0.008 in the radiation era.  

On the long strings in the network, if the functions ${\bf p}_+(u)$ and ${\bf p}_-(v)$ were smooth and varied on the scale of the correlation length (which is not far below the horizon scale), then the typical velocity of any crossing would be the inverse correlation length, and this would be the size of cusps.  What we see from Eq.~(\ref{V2}) is that such slow crossings and large cusps are impossible.  Superimposed on the slow motion is the irregular motion from the short distance structure, so all crossings occur at the much larger velocity $V \sim 0.1(G\mu)^{-(1-\chi)(1+2\chi)}  t^{-1}$, and all cusps have a size of order
\begin{equation}
V^{-1} \sim 10 (G\mu)^{(1-\chi)(1+2\chi)} t\ ,  \label{cusps}
\end{equation}
which is very much smaller than the correlation length.
Effectively each large cusp breaks up into a large number of small cusps: the high fractal dimension of ${\bf p}_+(u)$ and ${\bf p}_-(v)$ above the gravitational radiation length, noted in Ref.~\cite{Polchinski:2006ee}, implies that these curves will cross many times near the would-be large cusp.  We can estimate the number: the total length of each curve ${\bf p}_+(u), {\bf p}_-(v)$ during one Hubble time is $Vt \sim 0.1(G\mu)^{-(1-\chi)(1+2\chi)}$.  The total number of crossings is roughly the total length of either path divided by the solid angle $4\pi$ of the sphere, and so up to numerical factors it is large to the same extent that the size of each cusp is small.  We could also imagine course graining, looking at the string with a resolution $\delta\sigma$ that is longer than the gravitational radiation scale, and the same argument would imply $O((t/\delta\sigma)^{1-\chi})$ cusps each of size $O( t^{\chi} \delta\sigma^{1-\chi})$.

Recall further that the cusp region is the site of rapid loop production, so these small cusps will actually end up on small loops, disconnected from the long string.  It might seem paradoxical that the cusp size~(\ref{cusps}) is parametrically larger than the loop size~(\ref{lsize}), but this is another Lorentz effect: in the loop rest frame these sizes are of the same order.  

Regarding the large loops, we have already noted that these are likely to form without cusps.  However, when they eventually decay by gravitational radiation at $t \sim l/\Gamma G \mu$, the higher harmonics will decay more rapidly, smoothing out the string and likely leading to new cusps.  Thus there is the possibility of a cusp signal~\cite{Damour:2000wa,Siemens:2006vk} from these, although it will be reduced to the extent that the large loops represent only a fraction of the total loop length.

\section*{Acknowledgments}
 
JVR acknowledges financial support from {\it Funda\c{c}\~ao para a Ci\^encia e a Tecnologia}, Portugal, through grant SFRH/BD/12241/2003.  This work was supported in part by NSF grants PHY05-51164 and PHY04-56556.  We thank Lars Bildsten, Florian Dubath, Mark Hindmarsh, Vuk Mandic, Carlos Martins, Ken Olum, Paul Shellard, Xavier Siemens, and Alex Vilenkin for helpful discussions and communications.



\begin{thebibliography}{99}
\itemsep = 3pt



\bibitem{VilShell}
A. Vilenkin and E. Shellard, {\em Cosmic strings and other topological defects}
  (Cambridge Univ. Press 1994).

\bibitem{Hindmarsh:1994re}
M.~B. Hindmarsh and T.~W.~B. Kibble, Rept. Prog. Phys. {\bf 58},  477  (1995),
  hep-ph/9411342.

\bibitem{Bevis:2006mj}
N.~Bevis, M.~Hindmarsh, M.~Kunz and J.~Urrestilla,
  ``CMB power spectrum contribution from cosmic strings using  field-evolution
  simulations of the Abelian Higgs model,''
  Phys.\ Rev.\  D {\bf 75}, 065015 (2007)
  [arXiv:astro-ph/0605018].
  
\bibitem{Vanchurin:2005pa}
  V.~Vanchurin, K.~D.~Olum and A.~Vilenkin,
  ``Scaling of cosmic string loops,''
  Phys.\ Rev.\ D {\bf 74}, 063527 (2006)
  [arXiv:gr-qc/0511159].
  
\bibitem{Olum:2006ix}
  K.~D.~Olum and V.~Vanchurin,
  ``Cosmic string loops in the expanding universe,''
  arXiv:astro-ph/0610419.
  
\bibitem{Ringeval:2005kr}
  C.~Ringeval, M.~Sakellariadou and F.~Bouchet,
  ``Cosmological evolution of cosmic string loops,''
  JCAP {\bf 0702}, 023 (2007)
  [arXiv:astro-ph/0511646].
  
\bibitem{Martins:2005es}
  C.~J.~A.~Martins and E.~P.~S.~Shellard,
  ``Fractal properties and small-scale structure of cosmic string networks,''
  Phys.\ Rev.\ D {\bf 73}, 043515 (2006)
  [arXiv:astro-ph/0511792].

\bibitem{Polchinski:2006ee}
  J.~Polchinski and J.~V.~Rocha,
  ``Analytic study of small scale structure on cosmic strings,''
  Phys.\ Rev.\ D {\bf 74}, 083504 (2006)
  [arXiv:hep-ph/0606205].
  
\bibitem{Bennett:1989ak}
  D.~P.~Bennett and F.~R.~Bouchet,
  ``Cosmic string evolution,''
  Phys.\ Rev.\ Lett.\  {\bf 63}, 2776 (1989).
  
\bibitem{Quashnock:1990wv}
  J.~M.~Quashnock and D.~N.~Spergel,
  ``Gravitational self-interactions of cosmic strings,''
  Phys.\ Rev.\ D {\bf 42}, 2505 (1990).
  
\bibitem{Sakellariadou:1990ne}
  M.~Sakellariadou,
  ``Gravitational waves emitted from infinite strings,''
  Phys.\ Rev.\  D {\bf 42}, 354 (1990)
  [Erratum-ibid.\  D {\bf 43}, 4150 (1991)].
  
\bibitem{Hindmarsh:1990xi}
  M.~Hindmarsh,
  ``Gravitational radiation from kinky infinite strings,''
  Phys.\ Lett.\ B {\bf 251}, 28 (1990).


\bibitem{Siemens:2001dx}
  X.~Siemens and K.~D.~Olum,
  ``Gravitational radiation and the small-scale structure of cosmic  strings,''
  Nucl.\ Phys.\ B {\bf 611}, 125 (2001)
  [Erratum-ibid.\ B {\bf 645}, 367 (2002)]
  [arXiv:gr-qc/0104085].

\bibitem{Siemens:2002dj}
  X.~Siemens, K.~D.~Olum and A.~Vilenkin,
  ``On the size of the smallest scales in cosmic string networks,''
  Phys.\ Rev.\ D {\bf 66}, 043501 (2002)
  [arXiv:gr-qc/0203006].
  
\bibitem{Olum:2006at}
  K.~D.~Olum and A.~Vilenkin,
  ``Reionization from cosmic string loops,''
  Phys.\ Rev.\ D {\bf 74}, 063516 (2006)
  [arXiv:astro-ph/0605465].
  
\bibitem{Hogan:2006we}
  C.~J.~Hogan,
  ``Gravitational waves from light cosmic strings: Backgrounds and bursts with
  large loops,''
  Phys.\ Rev.\ D {\bf 74}, 043526 (2006)
  [arXiv:astro-ph/0605567].
  
\bibitem{Siemens:2006yp}
  X.~Siemens, V.~Mandic and J.~Creighton,
  ``Gravitational wave stochastic background from cosmic (super)strings,''
  arXiv:astro-ph/0610920.

\bibitem{wip} 
F. Dubath and J.~Rocha, 
``Periodic gravitational waves from small cosmic string loops,'' 
in preparation.

\bibitem{Pogosian:2006hg}
  L.~Pogosian, I.~Wasserman and M.~Wyman,
  ``On vector mode contribution to CMB temperature and polarization from  local
  strings,''
  arXiv:astro-ph/0604141;\\
  U.~Seljak and A.~Slosar,
  ``B polarization of cosmic microwave background as a tracer of strings,''
  Phys.\ Rev.\ D {\bf 74}, 063523 (2006)
  [arXiv:astro-ph/0604143].

\bibitem{Jenet:2006sv}
 F.~A.~Jenet {\it et al.},
  ``Upper bounds on the low-frequency stochastic gravitational wave  background
  from pulsar timing observations: Current limits and future  prospects,''
  Astrophys.\ J.\  {\bf 653}, 1571 (2006)
  [arXiv:astro-ph/0609013].
  
\bibitem{Albrecht:1989mk}
  A.~Albrecht and N.~Turok,
  ``Evolution of cosmic string networks,''
  Phys.\ Rev.\ D {\bf 40}, 973 (1989).

\bibitem{Damour:2000wa}
  T.~Damour and A.~Vilenkin,
  ``Gravitational wave bursts from cosmic strings,''
  Phys.\ Rev.\ Lett.\  {\bf 85}, 3761 (2000)
  [arXiv:gr-qc/0004075].

\bibitem{Siemens:2006vk}
  X.~Siemens, J.~Creighton, I.~Maor, S.~Ray Majumder, K.~Cannon and J.~Read,
  ``Gravitational wave bursts from cosmic (super)strings: Quantitative
  analysis and constraints,''
  Phys.\ Rev.\ D {\bf 73}, 105001 (2006)
  [arXiv:gr-qc/0603115].

  
\end{thebibliography}
\end{document}